\documentclass[preprint,authoryear,12pt]{elsarticle}

\usepackage{amsmath,amssymb}
\usepackage{graphicx}% Include figure files
\usepackage{dcolumn}% Align table columns on decimal point
\usepackage{bm}% bold math
\usepackage{mathrsfs}
\usepackage{url}

\journal{Pattern Recognition Letters}

\begin{document}

\begin{frontmatter}

\title{Characterization and Exploitation of Community Structure in Cover Song Networks}

\author[label1]{Joan Serr\`a\fnref{label4}}\ead{joan.serraj@upf.edu} 
\author[label2]{Massimiliano Zanin}\ead{mzanin@innaxis.org}
\author[label1]{Perfecto Herrera}\ead{perfecto.herrera@upf.edu}
\author[label1]{Xavier Serra}\ead{xavier.serra@upf.edu}

\address[label1]{Music Technology Group, Universitat Pompeu Fabra, Roc Boronat 138, 08018 Barcelona, Spain.}
\address[label2]{INNAXIS Foundation \& Research Institute, Vel\'azquez 157, 28002 Madrid, Spain.\\ Center for Biomedical Technology, Polytechnic University of Madrid, Campus Montegancedo, 28223 Pozuelo de Alarc\'on, Madrid, Spain.}
\fntext[label4]{Corresponding author. Phone +34 93 542 2864, fax +34 93 542 2517.}


\begin{abstract}
The use of community detection algorithms is explored within the framework of cover song identification, i.e.~the automatic detection of different audio renditions of the same underlying musical piece. Until now, this task has been posed as a typical query-by-example task, where one submits a query song and the system retrieves a list of possible matches ranked by their similarity to the query. In this work, we propose a new approach which uses song communities (clusters, groups) to provide more relevant answers to a given query. Starting from the output of a state-of-the-art system, songs are embedded in a complex weighted network whose links represent similarity (related musical content). Communities inside the network are then recognized as groups of covers and this information is used to enhance the results of the system. In particular, we show that this approach increases both the coherence and the accuracy of the system. Furthermore, we provide insight into the internal organization of individual cover song communities, showing that there is a tendency for the original song to be central within the community. We postulate that the methods and results presented here could be relevant to other query-by-example tasks.
\end{abstract}

\end{frontmatter}

\section{Introduction \label{sec:intro}}

Audio cover song identification is the task of automatically detecting which songs are versions of the same underlying musical piece using only information extracted from their raw audio signal \citep{Serra10BOOKCHAP}. This addresses an important problem faced by modern society: the classification and organization of digital information. More concretely, it addresses the detection of near-duplicate musical documents \citep{Casey08IEEE}.

Cover song identification is a challenging task, since cover songs might differ from their originals in several musical aspects such as timbre, tempo, song structure, main tonality, arrangement, lyrics, or language of the vocals \citep{Serra10BOOKCHAP}. Nevertheless, the identification of cover song versions has been a very active area of study within the music information retrieval (MIR) community over the last years \citep{Serra10BOOKCHAP, Casey08IEEE, Downie08AST}. Thanks to these efforts, and to the development of a number of specific tools to extract and analyze musical information from audio \citep{Casey08IEEE}, we now dispose of a variety of metrics for the estimation of the similarity between cover songs \citep{Serra10BOOKCHAP}.

These metrics are commonly used to search for covers in a music collection, ranking the relevance of each song to a given query. Indeed, cover song identification has been traditionally set up as a typical information retrieval (IR) task of query-by-example \citep{BaezaYates99BOOK, Manning08BOOK}, where the user submits a query (a song) and receives an answer back (a list of songs ranked by their relevance to the query). In the present article we propose a novel approach: after processing isolated queries through query-by-example, systems may focus on groups of items, with the new aim of identifying communities of songs within a given music collection\footnote{Through the manuscript we use the words group, set, community, or cluster interchangeably.}.

Using such a strategy has many intuitive advantages. Importantly, one should bear in mind that these advantages are not specific for the cover song detection task, and hold for any IR system operating through query-by-example \citep{BaezaYates99BOOK, Manning08BOOK}, including analogous systems such as recommendation systems \citep{Resnick97CACM}. First, given that current systems provide a suitable metric to quantify the similarity between query items, several well-researched options exist to exploit this information in order to detect inherent groups of items \citep{Xu09BOOK, Jain99ACM, Fortunato09BOOKCHAP, Danon05JSM}. Second, focusing on groups of items may help the system in retrieving more coherent answers for isolated queries. In particular, the answers to any query belonging to a given group would coherently contain the other songs in the group, an advantage that is not guaranteed by query-by-example systems alone. Third, music collections are usually organized and structured on multiple scales. Thus we can infer and exploit these regularities to increase the overall accuracy of traditional cover song identification systems. Note that the two previous advantages specifically aim to achieve higher user satisfaction and confidence in IR systems, as they can be perceived as rational agents or assistants. Finally, once groups of coherent items are correctly detected, one can study these groups in order to retrieve new information, either from the individual communities or from the relations between these.

In this article, for automatically identifying cover song sets (or groups) in a music collection we employ a number of unsupervised grouping algorithms on top of a state-of-the-art query-by-example system \citep{Serra09NJP}. We consider clustering algorithms \citep{Xu09BOOK, Jain99ACM} and, in particular, community detection algorithms \citep{Fortunato09BOOKCHAP, Danon05JSM}. The reader may easily see the resemblance between the detection of cover song sets and a more classical community detection task inside a complex network \citep{Boccaletti06PR, Costa08preprint}. This way, a set of nodes $\mathscr{N} \equiv \{n_1, n_2, \ldots, n_N \}$ represents the $N$ recordings being analyzed, and the elements of the $N \times N$ weight matrix $\mathscr{W}$ represent the distance (dissimilarity) between any couple of nodes. Provided that the weights of this matrix are assigned with the help of a suitable cover song dissimilarity metric (e.g.~the same one used to originally rank the answer to a query), communities inside this complex network will represent sets of recordings with related musical content. Although complex networks and community detection algorithms have been used in many problems involving complex systems \citep{Boccaletti06PR, Costa08preprint}, and more specifically in studying musical networks \citep{Buldu07NJP, Teitelbaum08C, Cano06C}, to the best of our knowledge they have never been applied in the context of a retrieval task before. The only exception is our previous work \citep{Serra09ISMIR}, of which the present article shows considerable extensions, improvements, and new results. An alternative technique for improving cover song retrieval was considered in \citep{Lagrange10ISMIR}.

We now provide a brief overview of our main contributions and, at the same time, outline the remaining structure of the article. We first build and analyze a cover song network (Sec.~\ref{sec:coversongnet}). To this end we apply a state-of-the-art algorithm for cover song similarity to an in-house music collection. We then do an analysis of this network, both of its topology and of the characteristics of the percolation process. Within this analysis we find a strong modular structure, with well-defined communities and a clustering coefficient higher than expected in an equivalent random network. This confirms our intuitive reasoning that cover songs naturally cluster into cover song sets. With this knowledge we can then safely proceed to detect the actual sets of covers based on the output of the state-of-the-art algorithm (Sec.~\ref{sec:SetDetection}). For that, several clustering and community detection strategies are compared. Four of these strategies are based on community detection in complex networks, of which three of them are novel contributions. An assessment of the computation time of all the considered methods is also done. Next, we show how query-by-example results can be improved by incorporating the information obtained through the group detection stage into the system (Sec.~\ref{sec:AccuracyImprovement}). Indeed, our results show a coherent increase in the accuracy of the system, with particularly promising values for community detection methods. This confirms our intuitive reasoning that exploiting the regularities found in the answers given by a query-by-example system can lead to an overall accuracy increase. Finally, we focus on the internal organization of cover song sets. More concretely, a pioneering study of the role that original songs (i.e.~the ones performed by the original author or artist) play within a group of covers is done (Sec.~\ref{sec:Originals}). To the authors' knowledge, the present study is the first attempt done in this direction. In particular, we show that there is a tendency for the original song to be central within the community. A short conclusions section closes the article (Sec.~\ref{sec:Conclusions}).

\section{Cover song networks \label{sec:coversongnet}}

\subsection{Building the network \label{sec:NetworkBuilding}}

The first step required by our proposal is to create a network and to embed nodes (songs) into it. We use an in-house music collection of 2125 songs comprising a variety of genres and styles. This collection is an extension of the one used by \citet{Serra09NJP}, to which we refer for further details, and consists of 523 non-overlapping groups of cover songs, each group having an identificatory label which we use in the evaluation stages. The cardinality of these groups, i.e.~the number of songs per group, varies between 2 and 18, with an expected value of 4.

Links between network nodes should represent the cover song relationship between corresponding musical pieces (the dissimilarity between their musical content). Therefore, an algorithm to compute this dissimilarity is needed in order to calculate the elements $w_{i,j}$ of the matrix $\mathscr{W}$ for each couple of nodes $n_i$ and $n_j$. Several alternatives for such dissimilarity measures have been proposed in the literature \citep{Serra10BOOKCHAP}. In particular, we use the $Q_{\max}$ measure presented by \citet{Serra09NJP}. This measure allows to track all potential differences between cover songs of the same underlying musical piece (Sec.~\ref{sec:intro}). However, in spite of being one of the most promising strategies proposed so far, its accuracy is not perfect. This is a further motivation to improve the accuracy of the system through a post-processing step based on cover set detection.

A brief outline of the $Q_{\max}$ measure follows. First, a time series of musical descriptors is extracted for all songs. In the case of cover songs, tonal similarity is commonly exploited \citep{Serra10BOOKCHAP}. In particular, $Q_{\max}$ employs time series of pitch class profiles \citep[PCP;][]{Gomez06THESIS}. PCP features estimate the amount of energy for each musical note of the Western musical scale that is present in a short analysis frame of the raw audio signal. This analysis is performed in a moving window, leading to a time series that is robust against non-tonal components (e.g. ambient noise or percussive sounds), and independent of timbre and the specific instruments used. Furthermore, PCPs are independent of a musical piece's loudness and volume fluctuations. As cover versions may be played in different tonalities (e.g.~to be adapted to the characteristics of a particular singer or instrument) one has to tackle differences in the main key of the song. This can be effectively done through various strategies \citep{Serra10BOOKCHAP}.

From the above PCP time series, one forms a state space representation for each song using delay coordinates \citep{Kantz04BOOK}. These representations are then compared on a pairwise basis through a cross recurrence plot (CRP), which is the bivariate generalization of classical recurrence plots \citep{Eckmann87EPL, Marwan07PR}. Finally, the $Q_{\max}$ measure is used to extract features that are sensitive to cover song CRP characteristics. This measure was derived from a previously published RQA measure [$L_{\max}$, \citet{Eckmann87EPL}], but adapted to the problem at hand by allowing to track curved and potentially disrupted traces in a CRP. Despite this adaptation, in \citet{Serra09NJP} we showed that the $Q_{\max}$ measure is not restricted to MIR nor to the particular application of cover song identification.

\begin{figure}[tb]
 \centerline{
  \includegraphics[width=1\linewidth]{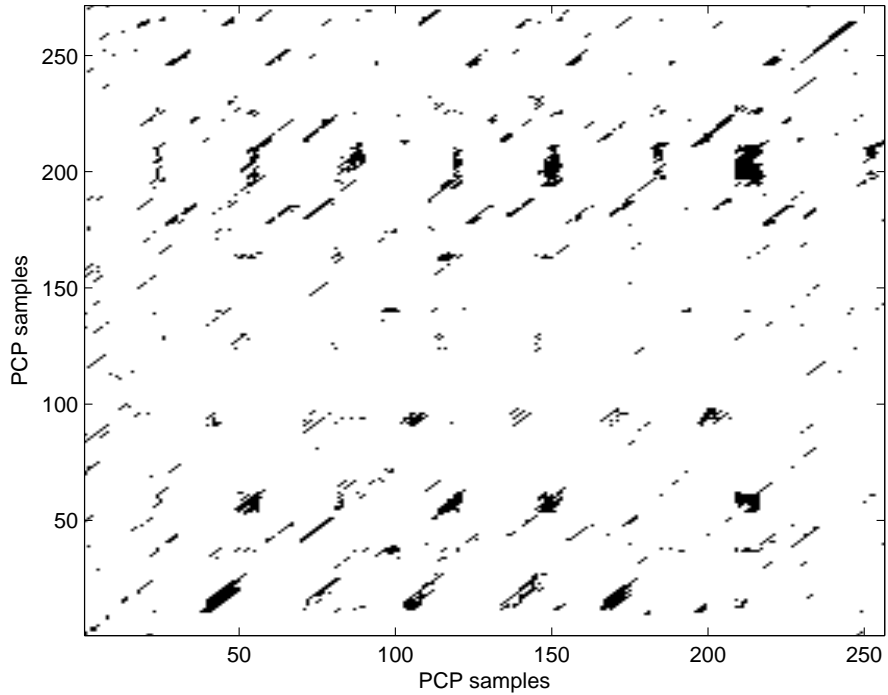}
 }
 \caption[sdh]{CRP for the song ``Rock around the clock'' as performed by Elvis Presley ($x$ axis) and The Sex Pistols ($y$ axis). Axes represent time and black dots represent correspondences between the tonal content of both songs. We see quite long black traces through the CRP, which are usually not straight diagonals but curved and disrupted ones, indicating similarly evolving temporal patterns in both song representations.}
 \label{fig:figExample1}
\end{figure}

\begin{figure}[tb]
\centerline{
  \includegraphics[width=1\linewidth]{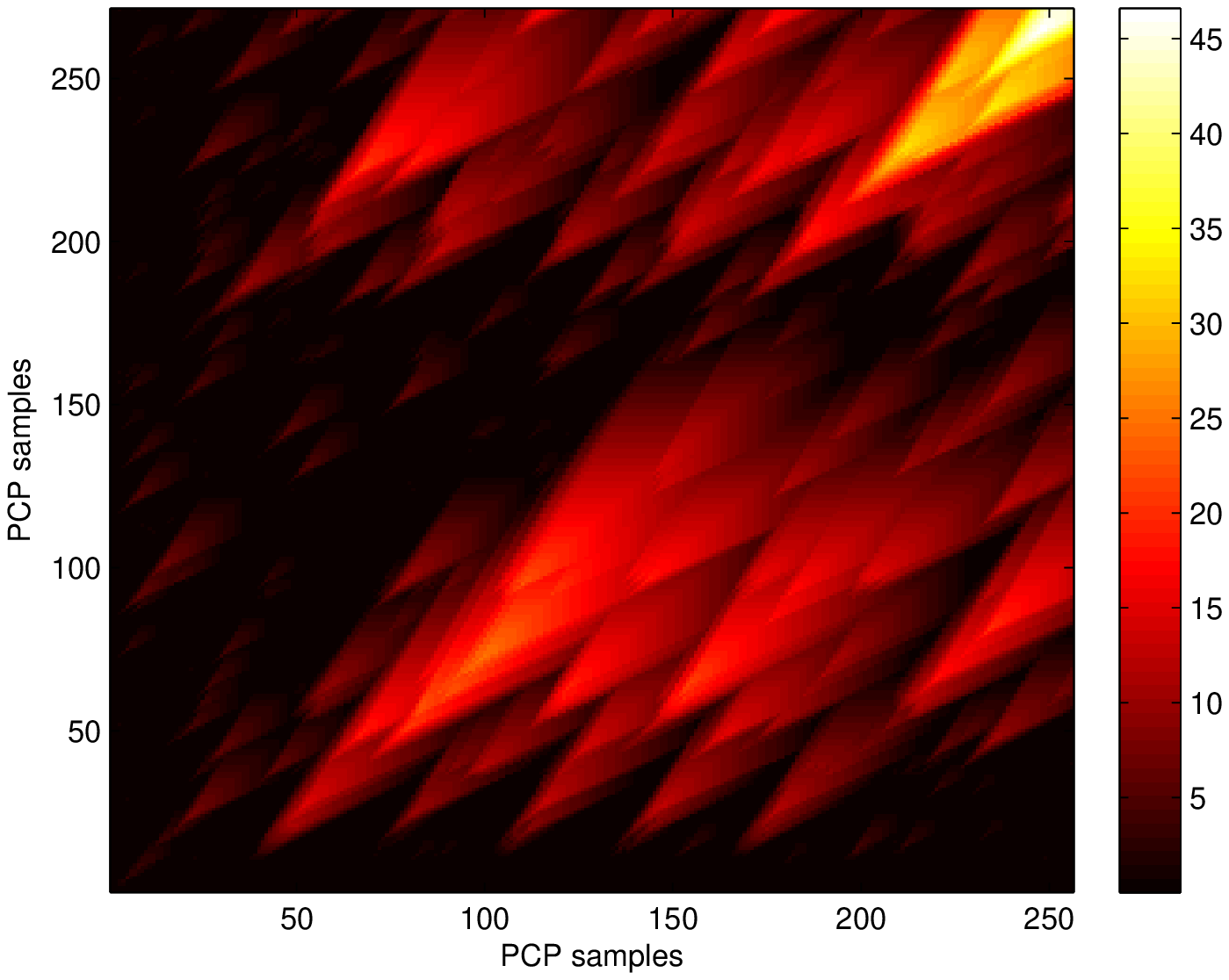}
 }
 \caption[sdh]{The $Q$ matrix \citep{Serra09NJP} for the same pair of songs. This matrix quantifies the lengths of the previously mentioned black traces. The $Q_{\max}$ measure corresponds to the maximum value in $Q$ (in this example $Q_{\max}=46.6$).}
 \label{fig:figExample2}
\end{figure}

An example of the abovementioned process is shown in Figs.~\ref{fig:figExample1} and~\ref{fig:figExample2}, which compares the song ``Rock around the clock'' as performed by Elvis Presley versus a version performed by The Sex Pistols. Since it is not the objective of this article to thoroughly present the $Q_{\max}$ measure, the interested reader is referred to \citet{Serra09NJP} for further details. A comprehensive overview of cover song similarity measures can be found in \citet{Serra10BOOKCHAP}.

The symmetric measure $Q_{\max}$ represents similarity: the higher the value, the more similar both analyzed recordings are in terms of their tonal musical content. To fill the weighted adjacency matrix $\mathscr{W}$ of the network, we proceed as in \citet{Serra09MIREX} and convert $Q_{\max}$ to a dissimilarity value by taking
\begin{equation}
w_{i,j}  = \frac{\sqrt{\left\vert s_j \right\vert}}{{Q_{\max } \left(s_i,s_j\right)}} ,
\end{equation}
where $\vert s_j \vert$ is proportional to the duration of song $s_j$ and $Q_{\max } \left(s_i,s_j\right)\in\left[1,\max\left(\vert s_i \vert,\vert s_j \vert\right)\right]$. Notice that $w_{i,j} = w_{j,i}$, iff $s_i$ and $s_j$ have the same duration. Recall that the nodes of the network $\mathscr{N} \equiv \{n_1, n_2, \ldots, n_N \}$ represent the $N$ recordings $s_i$ being analyzed.

\subsection{Analysis of the network \label{sec:NetworkAnalysis}}

\begin{figure*}[tb]
 \centerline{\includegraphics[width=0.9\linewidth]{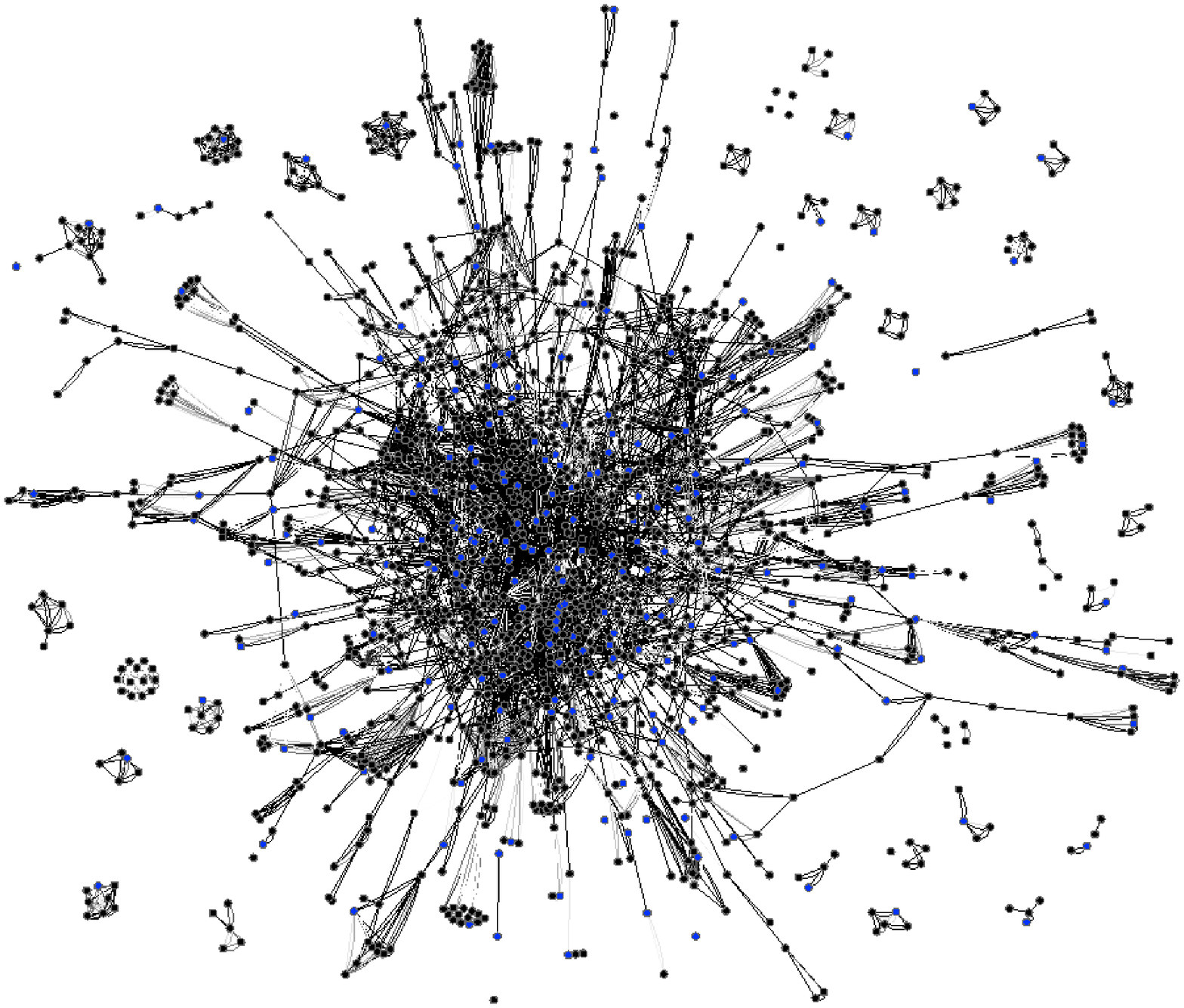}}
 \caption[sdh]{Graphical representation of the cover song network when a threshold of $0.2$ is applied. Original songs are drawn in blue, while covers are in black. In Sec.~\ref{sec:Originals}, the role of original songs inside each community will be further studied.}
 \label{fig:figNet}
\end{figure*}

The result of the previous procedure over the available data is a weighted directed graph expressing cover song relationships. This resulting network is represented in Fig.~\ref{fig:figNet}. A threshold has been applied so that only pairs of nodes with $w_{i,j} \leq 0.2$ are drawn. Some clusters, that is, sets of covers, are already visible, especially in the external zones of the network.

\begin{figure*}[!tb]
 \begin{center}
  \includegraphics[width=0.48\linewidth]{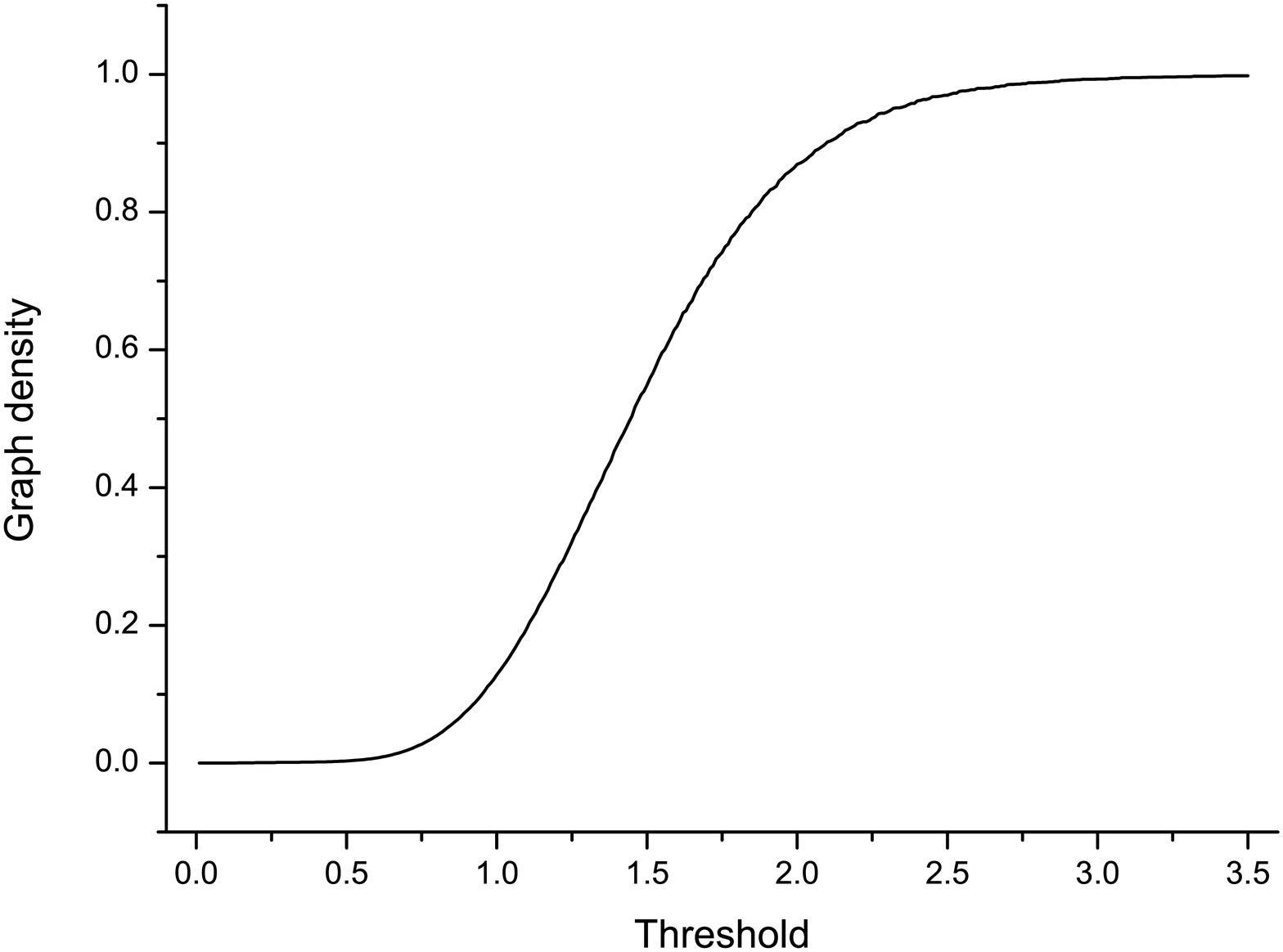}
  \includegraphics[width=0.48\linewidth]{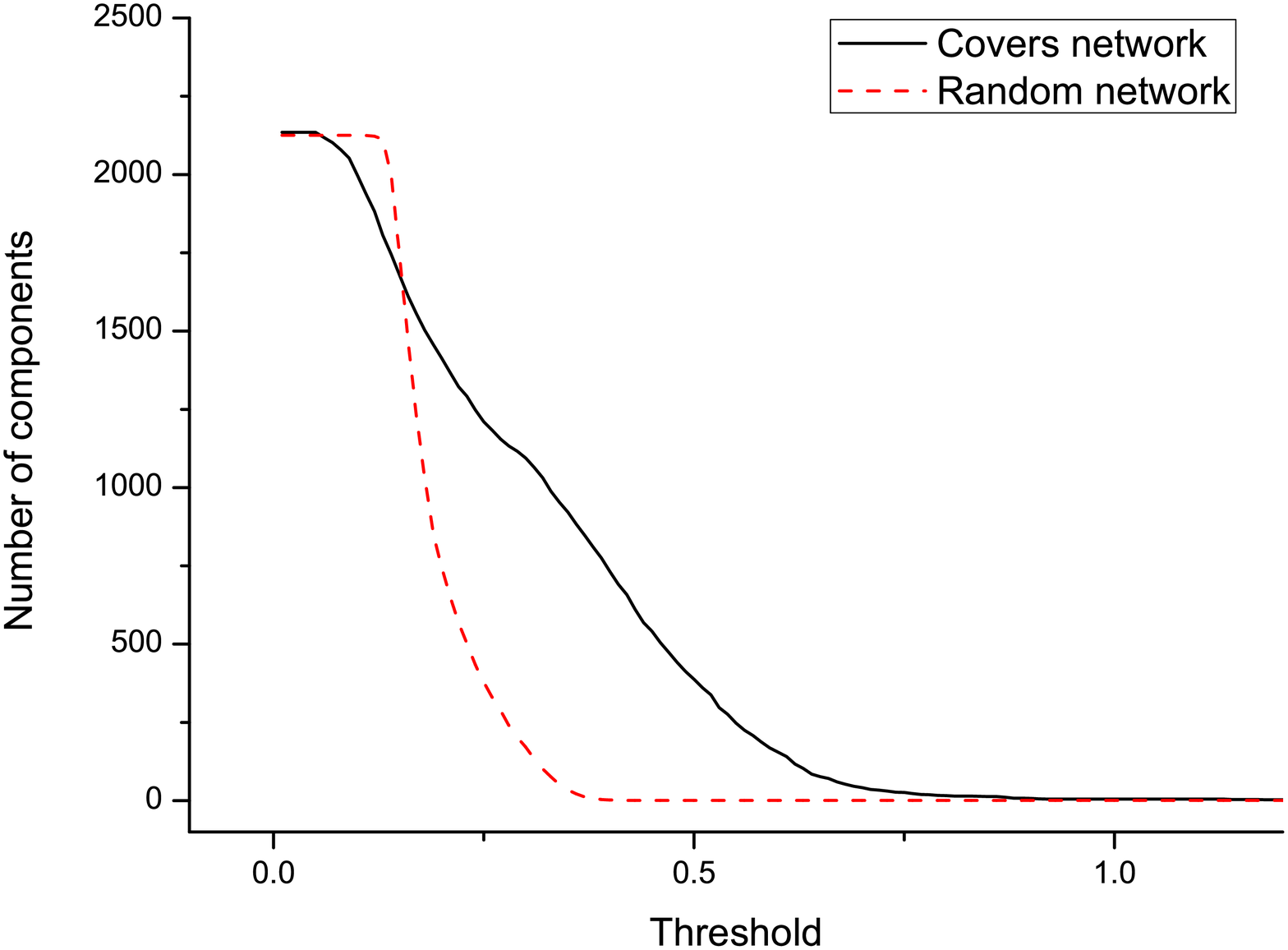} \\
  \includegraphics[width=0.48\linewidth]{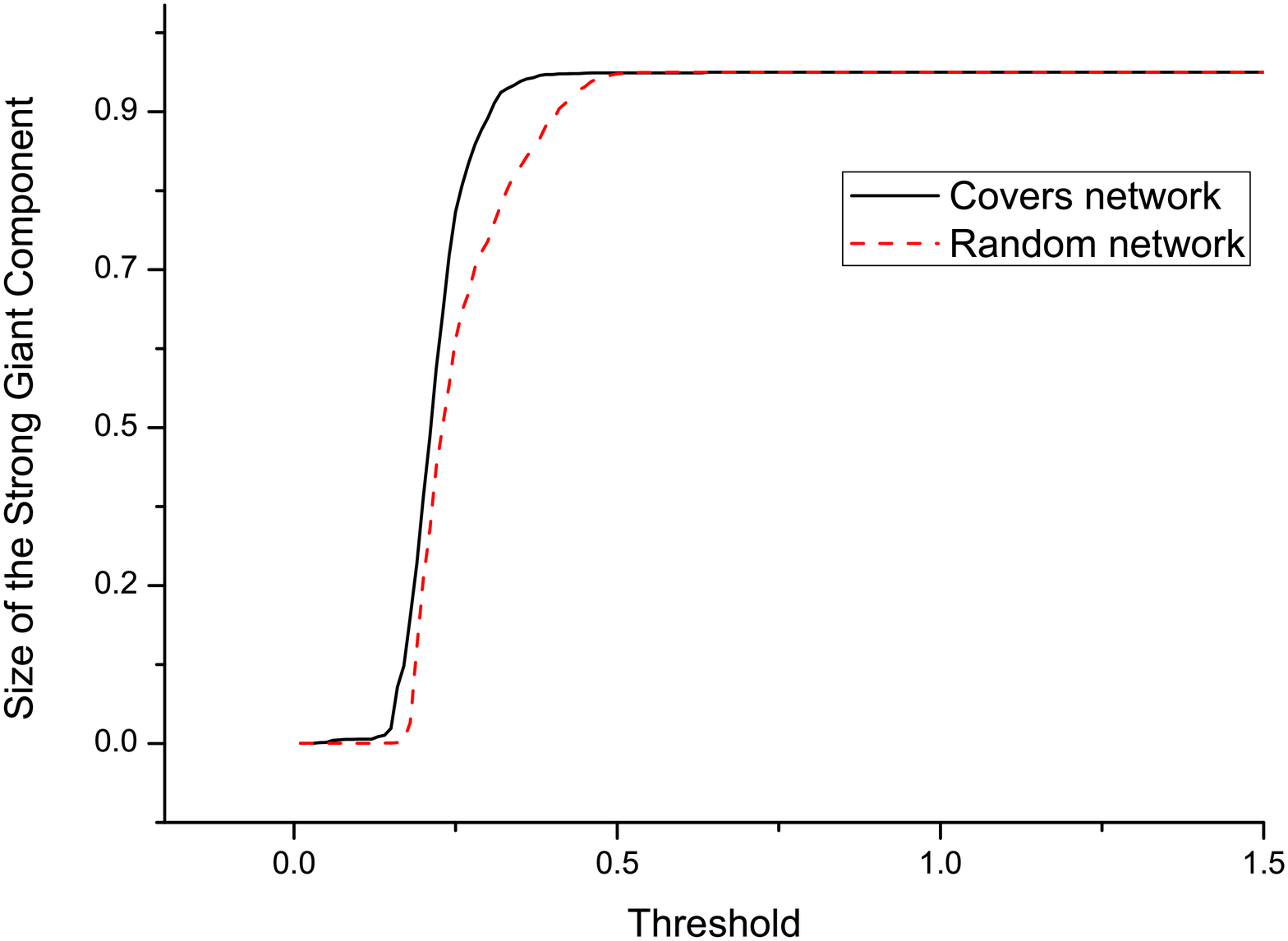}
  \includegraphics[width=0.48\linewidth]{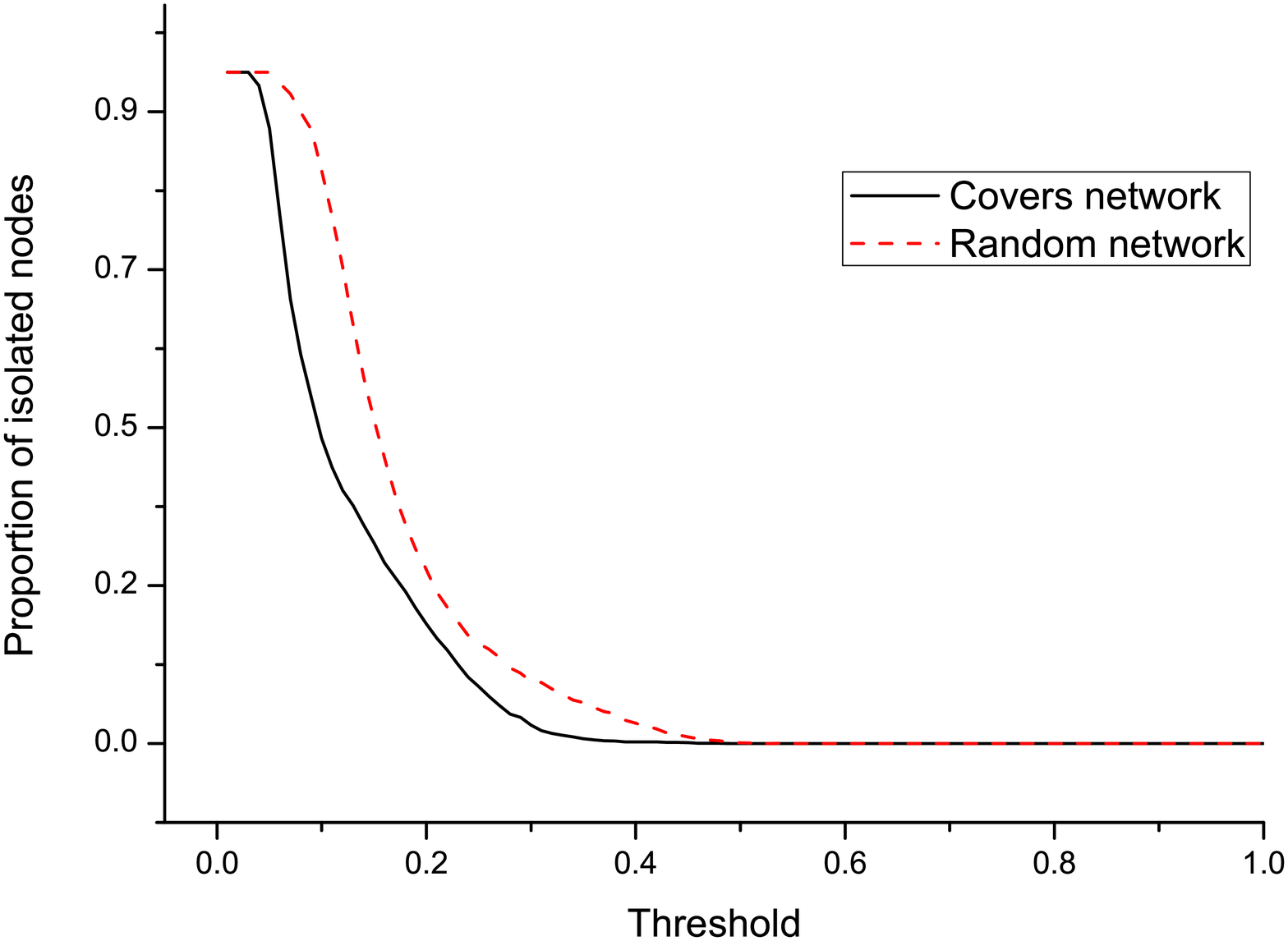} \\
  \includegraphics[width=0.48\linewidth]{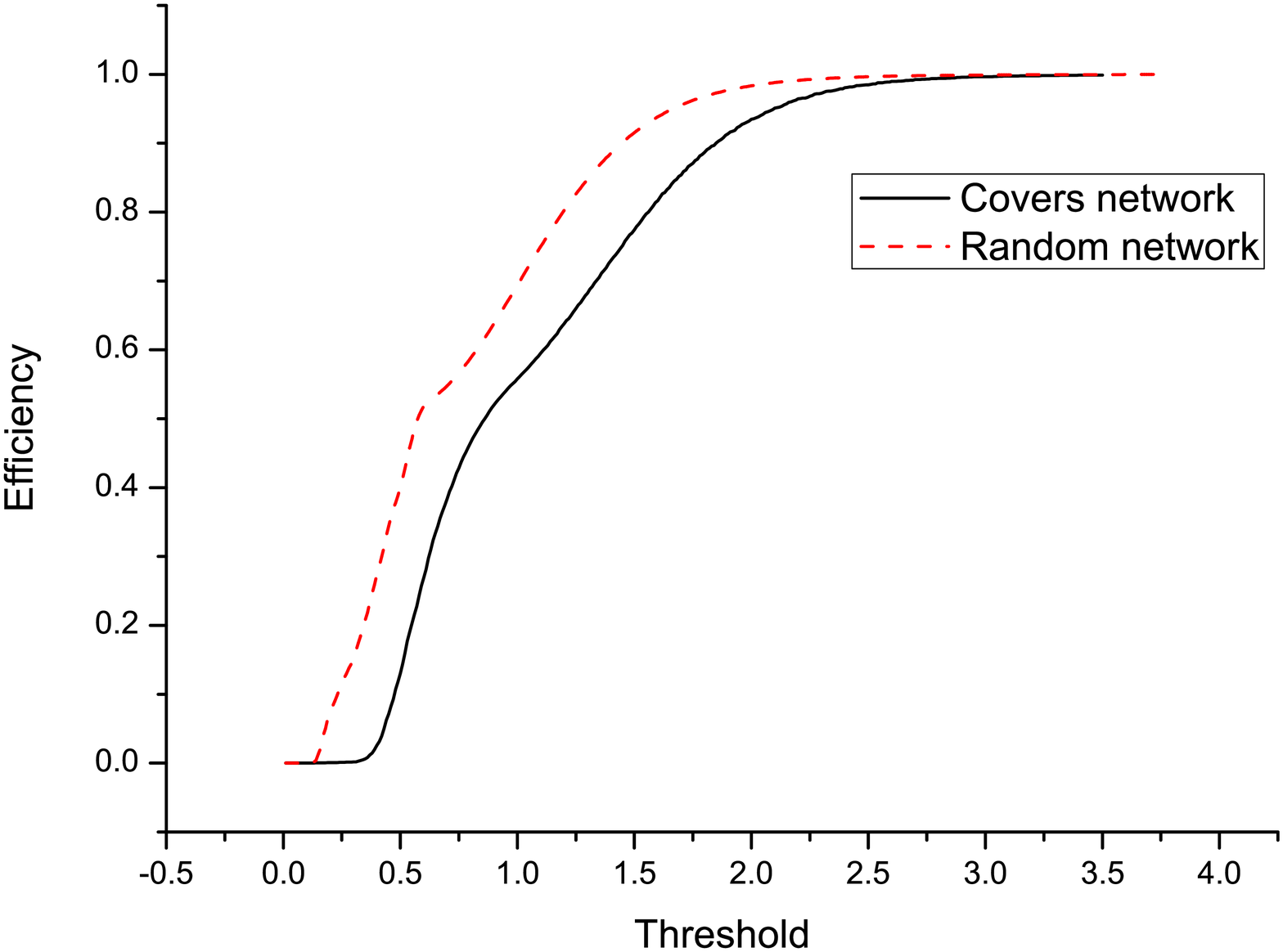}
  \includegraphics[width=0.48\linewidth]{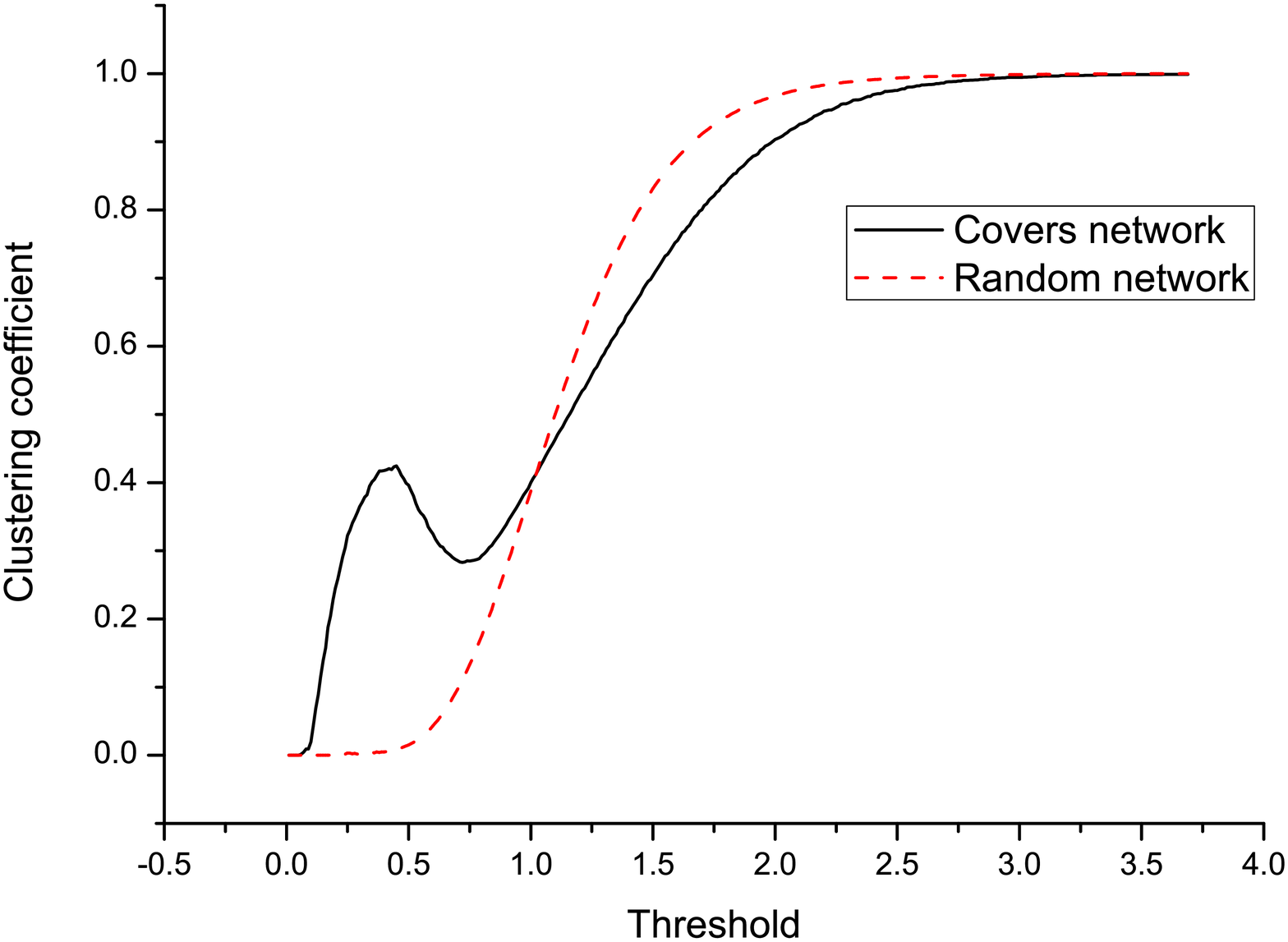} \\
 \end{center}
 \caption[sdh]{(Black solid lines) Evolution of six metrics of the network as a function of the threshold. These metrics are, from top left to bottom right: graph density, number of independent components, size of the strong giant component, number of isolated nodes, efficiency, and clustering coefficient. (Red dashed lines) Expected value in a random network with the same number of nodes and links. Nearly identical figures were obtained when considering a symmetric dissimilarity matrix (see text).}
 \label{fig:figMetrics}
\end{figure*}

In order to understand how the network evolves when the threshold is modified, we represent six different classical network metrics as a function of the threshold (Fig.~\ref{fig:figMetrics}). These metrics correspond to \citep{Boccaletti06PR}: graph density, number of independent components, size of the strong giant component, number of isolated nodes, efficiency \citep{Latora01PRL}, and clustering coefficient. In the same plots, we also display the values for the last five measures as expected in random networks with the same number of nodes and links.

By looking at the evolution of these metrics, we can infer some interesting knowledge about the network and its inherent structure. Notice that when reducing the threshold (and therefore increasing the deleted links), the network splits into a higher number of clusters than expected (Fig.~\ref{fig:figMetrics}, top right), which represents the formation of cover song communities. This process begins around a threshold of $0.5$ (see, for instance, the evolution of the size of the strong giant component). When these communities are formed, they maintain a high clustering coefficient and a high triangular coherence (bottom right graph of Fig.~\ref{fig:figMetrics}, between $0.3$ and $0.5$), i.e.~sub-networks of covers tend to be fully connected. It is also interesting to note that the number of isolated nodes remains lower than expected, except for high thresholds (Fig.~\ref{fig:figMetrics}, middle right). This suggests that most of the songs are connected to some cluster while a small group of them are different, with unique musical features. We found nearly identical results using a symmetric dissimilarity matrix $\mathscr{W}'$ with $w'_{i,j}=w'_{j,i}=\left(w_{i,j}+w_{j,i}\right)/2$.

\section{Detecting groups of covers \label{sec:SetDetection}}

We assess the detection of cover sets (or communities) by evaluating a number of unsupervised methods either based on clustering or on complex networks. Three of these are novel approaches. Since standard implementations of clustering algorithms do not operate with an asymmetric dissimilarity measure, in this section and in the subsequent one we use the symmetric dissimilarity matrix $\mathscr{W}'$ explained above.

\subsection{Methods \label{sec:Methods1}}

\begin{description}

\item[K-medoids] K-medoids (KM) is a classical technique to group a set of objects inside a previously known number of $K$ clusters. This algorithm is a common choice when the computation of means is unavailable (as it solely operates on pairwise distances) and can exhibit some advantages compared to the standard K-means algorithm \citep{Xu09BOOK}, in particular with noisy samples. The main drawback for its application is that, as well as with the K-means algorithm, the K-medoids algorithm needs to set $K$, the number of expected clusters. However, several heuristics can be used for that purpose. We employ the K-medoids implementation of the \textit{tamo} package\footnote{\url{http://fraenkel.mit.edu/TAMO}}, which incorporates several heuristics to achieve an optimal $K$ value%\footnote{\url{http://fraenkel.mit.edu/TAMO/documentation/TAMO.Clustering.Kmedoids.html}}
. We use the default parameters and try all possible heuristics provided in the implementation.

\item[Hierarchical clustering] Four representative agglomerative hierarchical clustering methods have been tested \citep{Xu09BOOK, Jain99ACM}: single linkage (SL), complete linkage (CL), group average linkage (UPGMA), and weighted average linkage (WPGMA). We use the \textit{hcluster} implementation\footnote{\url{http://code.google.com/p/scipy-cluster}} with the default parameters, and we try different cluster validity criteria such as checking descendants for inconsistent values, or considering the maximal or the average inter-cluster cophenetic distance%\footnote{\url{http://www.soe.ucsc.edu/~eads/cluster.html}}
. Thus, in the end, all clustering algorithms rely only on the definition of a distance threshold $d'_{\mathrm{Th}}$, which is set experimentally.

\item[Modularity optimization] This method (MO), as well as the next three algorithms, is designed to exploit a complex network collaborative approach. MO extracts the community structure from large networks based on the optimization of the network modularity \citep{Fortunato09BOOKCHAP, Danon05JSM}. In particular, we use the method proposed in \citet{Blondel08JSM} with the implementation by Aynaud\footnote{\url{http://perso.crans.org/~aynaud/communities/index.html}}. This method is reported to outperform all other known community detection algorithms in terms of computational time while still maintaining a high accuracy.

\item[Proposed method 1] Our first proposed method (PM1) applies a threshold to each network link in order to create an unweighted network where two nodes are connected only if their weight (dissimilarity) is less than a certain value $w'_{\mathrm{Th}}$. In addition, for each row of $\mathscr{W}'$, we only allow a maximum number of connections, considering only the lowest values of the thresholded row as valid links. That is, we only consider the first $r'_{\mathrm{Th}}$ nearest neighbors for each node (values $w'_{\mathrm{Th}}$ and $r'_{\mathrm{Th}}$ are set experimentally). Finally, each connected component is assigned to be a group of covers. Although this is a very na\"ive approach, it will be shown that, given the considered network and dissimilarity measure, it achieves a high accuracy level at low computational costs.

\item[Proposed method 2] The previous approach could be further improved by reinforcing triangular connections in the complex network before the last step of checking for connected components. In other words, proposed method 2 (PM2) tries to reduce the ``uncertainty'' generated by triplets of nodes connected by two edges and to reinforce coherence in a triangular sense.

This idea can be illustrated by the following example (Fig.~\ref{fig:figPM2}). Suppose that three nodes in the network, e.g.~$n_i$, $n_j$, and $n_k$, are covers: the resulting subnetwork should be triangular, so that every node is connected with the two remaining ones. On the other hand, if $n_i$, $n_j$, and $n_k$ are not covers, no edge should exist between them. If couples $n_i, n_j$ and $n_i, n_k$ are respectively connected (Fig.~\ref{fig:figPM2}A), we can induce more coherence by either deleting one of the existing edges (Fig.~\ref{fig:figPM2}B), or by creating a connection between $n_j$ and $n_k$ (i.e.~forcing the existence of a triangle, Fig.~\ref{fig:figPM2}C). This coherence can be measured through an objective function $f_{\text{O}}$ which considers complete and incomplete triangles in the whole graph. We define $f_{\text{O}}$ as a weighted difference between the number of complete triangles $N_\bigtriangledown$ and the number of incomplete triangles $N_{\vee}$ (three vertices connected by only two links) that can be computed from a pair of vertices: $f_{\text{O}}(N_\bigtriangledown,N_{\vee}) = N_\bigtriangledown - \alpha N_{\vee}$. The constant $\alpha$, which weights the penalization for having incomplete triangles, is set experimentally.
\begin{figure}[tb]
 \centerline{\includegraphics[width=0.7\linewidth]{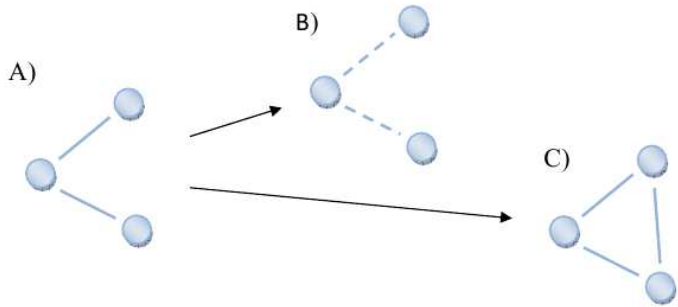}}
 \caption[sdh]{Example of the process of reinforcing the triangular coherence of the network. The sub-network in the left part (A) can be improved by either deleting a link (B), or by adding a third link between the two nodes that were not originally connected (C).} \label{fig:figPM2}
\end{figure}

The implementation of this idea sequentially analyzes each pair of vertices $n_i,n_j$ by calculating the value of $f_{\text{O}}$ for two situations: (i) when an edge between $n_i$ and $n_j$ is artificially created and (ii) when such an edge is deleted. Then, the option which maximizes $f_{\text{O}}$ is kept and the adjacency matrix is updated as necessary. The process of assigning cover sets is the same as in PM1.

\item[Proposed method 3] The computation time of the previous method can be substantially reduced by considering for the computation of $f_{\text{O}}$ only those vertices whose connections seem to be uncertain. This is what proposed method 3 (PM3) does: if the dissimilarity between two songs is extremely high or low, this means that the cover song identification system has clearly detected a match or a mismatch. Accordingly, we only consider for $f_{\text{O}}$ the pairs of vertices whose edge weight is close to $w'_{\mathrm{Th}}$ (a closeness margin is empirically set).

\end{description}

\subsection{Evaluation methodology \label{sec:Methodology1}}

The experimental setup is an important aspect to be considered when evaluating cover song identification systems. Each setup is defined by different parameters \citep{Serra09ISMIR}: the total number of songs $N$, the number of cover sets $N_{\mathrm{C}}$ the collection includes, the cardinality $C$ of the cover sets (i.e.~the number of songs in the set), and the number of added noise songs $N_{\mathrm{N}}$ (i.e.~songs that do not belong to any cover set, which are included to add difficulty to the task). Because some setups can lead to wrong accuracy estimations \citep{Serra10BOOKCHAP}, it is safer to consider several of them, including fixed and variable cardinalities. In our experiments we use the setups summarized in Table \ref{table:Setups}. The whole network analyzed in Sec.~\ref{sec:NetworkAnalysis} corresponds to setup 3. For other setups we randomly sample cover sets from setup 3 and repeat the experiments $N_{\mathrm{T}}$ times. We either sample cover sets with a fixed cardinality ($C = 4$, the expected cardinality of setup 3) or without fixing it (variable cardinality, $C = \nu$). For sampled setups, the average accuracies reported.
\begin{table}[tb]
 \centering
 \begin{tabular}{lccccc}
  \hline
  Setup & \multicolumn{5}{c}{Parameters} \\
  \cline{2-6}
        & $N_{\mathrm{C}}$ & $C$ & $N_{\mathrm{N}}$ & $N$ & $N_{\mathrm{T}}$ \\
  \hline
  1.1 & 25 & 4 & 0 & 100 & 20 \\
  1.2 & 25 & $\nu$ & 0 & $\left\langle 100\right\rangle$ & 20 \\
  1.3 & 25 & 4 & 100 & 200 & 20 \\
  1.4 & 25 & $\nu$ & 100 & $\left\langle 200\right\rangle$ & 20 \\
  2.1 & 125 & 4 & 0 & 500 & 20 \\
  2.2 & 125 & $\nu$ & 0 & $\left\langle 500\right\rangle$ & 20 \\
  2.3 & 125 & 4 & 400 & 900 & 20 \\
  2.4 & 125 & $\nu$ & 400 & $\left\langle 900\right\rangle$ & 20 \\
  3 & 523 & $\nu$ & 0 & 2125 & 1 \\
  \hline
 \end{tabular}
 \caption{Experimental setup summary. The $\left\langle \cdot \right\rangle$ delimiters denote expected value.}
 \label{table:Setups}
\end{table}

To quantitatively evaluate cover set (or community) detection we use to the classical F-measure with even weighting \citep{BaezaYates99BOOK, Manning08BOOK},
\begin{equation}
	F = \frac{2 \bar{P} \bar{R}}{\bar{P} + \bar{R}},
	\label{eqn:F}
\end{equation}
which goes from 0 (worst case) to 1 (best case). In Eq.~(\ref{eqn:F}), $\bar{P}$ and $\bar{R}$ correspond to precision and recall, respectively. For our evaluation, we compute these two quantities independently for all songs and average afterwards, i.e.~unlike other clustering evaluation measures \citep{Sahoo06IKM}, $F$ is not computed on a per-cluster basis, but on a per-song basis. This way, and in contrast to the typical clustering F-measure or other clustering evaluation measures like Purity, Entropy, or F-Score \citep{Sahoo06IKM, Zhao02KDD}, we do not have to blindly choose which cluster is the representative for a given cover set. 

For each song $s_i$, we count the number of true positives $\tau_i^+$ (i.e.~the number of actual cover songs of $s_i$ estimated to belong to the the same community as $s_i$), the number of false positives $\tau_i^-$ (i.e.~the number of songs estimated to belong to the same group as $s_i$ that are actually not covers of $s_i$) and the number of false negatives $\upsilon_i^-$ (i.e.~the number of actual covers of $s_i$ that are not detected to belong to the same group as $s_i$). Then we define 
\begin{equation}
	P_i\! =\! \frac{\tau_i^+}{\tau_i^+ + \tau_i^-}
	\label{eqn:Pi}
\end{equation}
and
\begin{equation}
	R_i\! =\! \frac{\tau_i^+}{\tau_i^+ + \upsilon_i^-} .
	\label{eqn:Ri}
\end{equation}
These two quantities [Eqs.~(\ref{eqn:Pi}) and (\ref{eqn:Ri})] are averaged across all $N$ songs ($i=1,\ldots N$) to obtain $\bar{P}$ and $\bar{R}$, respectively.

\subsection{Results \label{sec:Results1}}

To assess the algorithms' accuracy we independently optimized all possible parameters for each algorithm. This optimization was done in-sample by a grid search, trying to maximize $F$ on the randomly chosen songs of setups 1.1 to 1.4. Within this optimization phase, we saw that the definition of a threshold (either $d'_{\mathrm{Th}}$ for clustering algorithms or $w'_{\mathrm{Th}}$ for community detection algorithms) was, in general, the only critical parameter for all algorithms (for our proposed methods we used $r'_{\mathrm{Th}}$ between 1 and 3). All other parameters turned out not to be critical for obtaining near-optimal accuracies. Methods that had specially broad ranges of these near-optimal accuracies were KM, PM2, and all considered hierarchical clustering algorithms.

We report the out-of-sample accuracies $F$ for setups 2.1 to 3 in Table~\ref{table:ResultsF}. Overall, the high $F$ values obtained (above 0.8 in the majority of the cases, some of them nearly reaching 0.9) indicate that the considered approaches are able to effectively detect groups of cover songs. This allows the possibility to reinforce the coherence within answers and to enhance the answer of a query-based retrieval system (see Sec.~\ref{sec:AccuracyImprovement}). In particular, we see that accuracies for PM1 and PM3 are comparable to the ones achieved by the other algorithms and, in some setups, even better. We also see that KM and PM2 perform worst.
\begin{table}[tb]
 \centering
 \begin{tabular}{lccccc}
  \hline
  Algorithm & \multicolumn{5}{c}{Setup} \\
  \cline{2-6}
            & 2.1 & 2.2 & 2.3 & 2.4 & 3\\
  \hline
  KM & 0.66 & 0.66 & 0.68 & 0.69 & \textit{n.c.} \\
  SL & 0.79 & 0.81 & 0.88 & 0.89 & 0.78 \\
  CL & 0.81 & 0.82 & 0.83 & 0.83 & 0.79 \\
  UPGMA & 0.82 & 0.83 & 0.83 & 0.83 & 0.79 \\
  WPGMA & 0.83 & 0.84 & 0.84 & 0.84 & 0.82 \\
  MO & 0.80 & 0.83 & 0.89 & 0.89 & 0.81 \\
  PM1 & 0.81 & 0.83 & 0.88 & 0.89 & 0.81 \\
  PM2 & 0.77 & 0.77 & \textit{n.c.} & \textit{n.c.} & \textit{n.c.} \\
  PM3 & 0.79 & 0.79 & 0.87 & 0.88 & 0.76 \\
  \hline
 \end{tabular}
 \caption{Accuracy $F$ for the considered algorithms and setups (see Table~\ref{table:Setups} for the details on the different setups). Due to algorithms' complexity, some results were not computed (denoted as \textit{n.c.}).}
 \label{table:ResultsF}
\end{table}
% \begin{table*}[tb]
%  \centering
%  \begin{tabular}{lccccc}
%   \hline
%   Algorithm & \multicolumn{5}{c}{Setup} \\
%   \cline{2-6}
%             & 2.1 & 2.2 & 2.3 & 2.4 & 3\\
%   \hline
%   KM & 0.657 & 0.662 & 0.681 & 0.692 & \textit{n.c.} \\
%   SL & 0.786 & 0.808 & 0.876 & 0.889 & 0.777 \\
%   CL & 0.811 & 0.817 & 0.829 & 0.826 & 0.791 \\
%   UPGMA & \textbf{0.823} & 0.827 & 0.829 & 0.826 & 0.791 \\
%   WPGMA & \textbf{0.825} & \textbf{0.842} & 0.844 & 0.843 & \textbf{0.815} \\
%   MO & 0.802 & 0.829 & \textbf{0.885} & \textbf{0.894} & \textbf{0.808} \\
%   PM1 & 0.807 & \textbf{0.834} & \textbf{0.881} & \textbf{0.890} & 0.807 \\
%   PM2 & 0.773 & 0.771 & \textit{n.c.} & \textit{n.c.} & \textit{n.c.} \\
%   PM3 & 0.787 & 0.786 & 0.865 & 0.876 & 0.763 \\
%   \hline
%  \end{tabular}
%  \caption{Accuracy $F$ for the considered algorithms and setups (see Table~\ref{table:Setups} for the details on the different setups). Due to algorithms complexity, some results were not computed (denoted as \textit{n.c.}). The two highest $F$ values for each setup are highlighted in bold.}
%  \label{table:ResultsF}
% \end{table*}

\subsection{Computation time \label{sec:Time}}

\begin{figure}[tb]
 \centering{\includegraphics[width=1\linewidth]{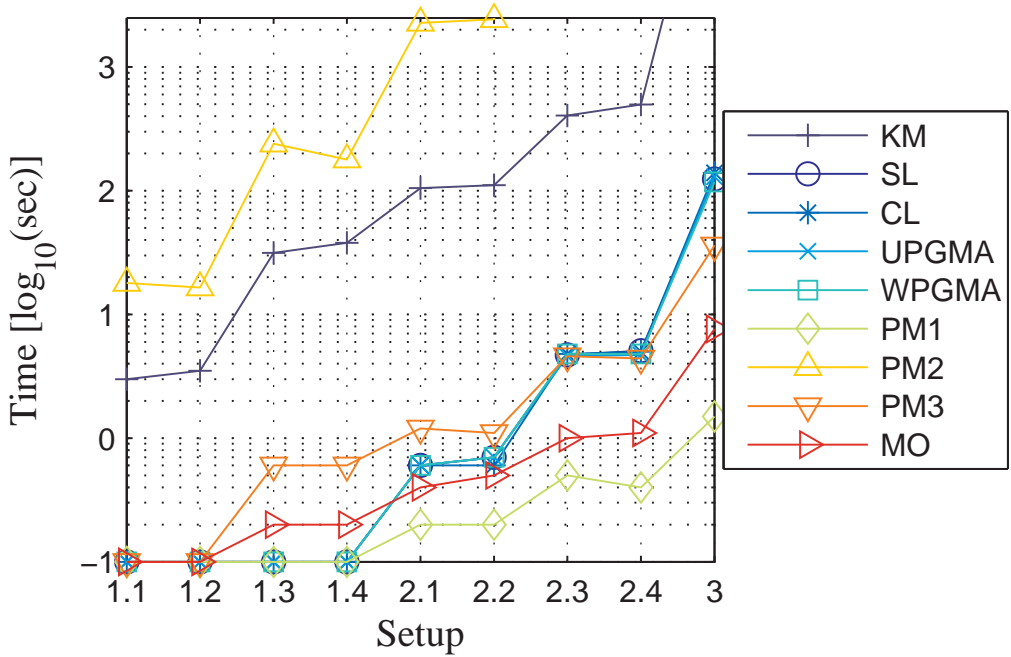}}
 \caption{Average time performance for each considered setup. Algorithms were run with an Intel(R) Pentium(R) 4 CPU 2.40GHz with 512M RAM.}
 \label{fig:Results_Time}
\end{figure}

In the application of these techniques to big real-world music collections, computational complexity is of great importance. To qualitatively evaluate this aspect, we report the average  amount of time spent by the algorithms to achieve a solution for each setup (Fig.~\ref{fig:Results_Time}). We see that KM and PM2 are completely inadequate for processing collections with more than 2000 songs (e.g.~setup 3). The steep rise in the time spent by hierarchical clustering algorithms to find a cluster solution for setup 3 also raises some doubts as to the usefulness of these algorithms for huge music collections [$O(N^2\log N)$, \citet{Jain99ACM}]. Furthermore, hierarchical clustering algorithms, as well as the KM algorithm, take the full pairwise dissimilarity matrix as input. Therefore, with a music collection of, say, 10 million songs, this distance matrix might be difficult to handle.

In contrast, algorithms based on complex networks show a better performance (with the aforementioned exception of PM2). More specifically, MO, PM1, and PM3 use local information (the nearest neighbors of the queries), while PM3 furthermore acts on a small subset of the links. It should also be noticed that the resulting network is very sparse, i.e.~the number of links is much lower than $N^2$ \citep{Boccaletti06PR} and, therefore, calculations on such graphs can be strongly optimized both in memory requirements and computational costs [as demonstrated, for instance, by \citet{Blondel08JSM}, who have applied their method to networks of millions of nodes and links].% Thus, methods based on complex networks represent a promising option for processing big music collections.

\section{Improving the accuracy through community detection \label{sec:AccuracyImprovement}}

In this section we investigate the use of the information obtained through the detection of communities to increase the overall accuracy of a query system.

\subsection{Method \label{sec:Method2}}

Given the dissimilarity matrix $\mathscr{W'}$ and a solution for the cluster or community detection problem, one can calculate a refined dissimilarity matrix $\mathscr{\hat{W}}$ by setting
\begin{equation}
	\hat{w}_{i,j} = \frac{w'_{i,j}}{\max(\mathscr{W'})} +\beta_{i,j} ,
	\label{eqn:refineW}
\end{equation}
where $\beta_{i,j}=0$ if $s_i$ and $s_j$ are estimated to be in the same community and $\beta_{i,j}=c$ otherwise. For ensuring songs in the same community to have $\hat{w}_{i,j}\leq 1$ and others to have $\hat{w}_{i,j}>1$, we use a constant $c>1$. This refined matrix $\mathscr{\hat{W}}$ can be used again to rank query answers according to cover song similarity and consequently, when compared to the initial $\mathscr{W}'$ of the original system, to evaluate the accuracy increase obtained.

\subsection{Evaluation methodology \label{sec:Methodology2}}

A common measure to evaluate query-by-example systems is the mean of average precisions (MAP) over all queries \citep{Manning08BOOK}, which we denote as $\left\langle \overline{P} \right\rangle$. To calculate such a measure, one averages across each of the answers $A_i$ to queries $s_i$, $A_i$ being an ascendingly ordered list according to the rows of $\mathscr{W}'$ (or $\mathscr{\hat{W}}$, depending on which solution we evaluate). More concretely, the average precision $\overline{P}_i$ for a query song $s_i$ is calculated from the retrieved answer $A_i$ as
\begin{equation}
	\displaystyle
	\overline{P}_i = \frac{1}{C-1} \sum_{r=1}^{N-1} P_i(r) I_i(r),
	\label{eqn:AP}
\end{equation}
where $P_i$ is the precision of the sorted list $A_i$ at rank $r$,
\begin{equation}
	\displaystyle
	P_i(r) = \frac{1}{r} \sum_{l=1}^{r} I_i(l),
	\label{eqn:AP_P}
\end{equation}
and $I_i$ is a relevance function such that $I_i(z)\! =\! 1$ if the song with rank $z$ in $A_i$ is a cover of $s_i$, $I_i(z)\! =\! 0$ otherwise. We then define the relative MAP increase as
\begin{equation}
	\Delta = 100 \left( \frac{\left\langle \overline{P}(\mathscr{\hat{W}}) \right\rangle}{\left\langle \overline{P}(\mathscr{W}') \right\rangle} -1 \right) .
	\label{eqn:relMAPincr}
\end{equation}
%where $\left\langle \overline{P} \right\rangle \in [0,1]$, although in our experiments,  $\left\langle \overline{P} \right\rangle$ never reached a value of 0.

\subsection{Results \label{sec:Results2}}

To assess the algorithms' accuracy we independently optimized the parameters for each algorithm as explained in the previous section. However, we now try to maximize $\left\langle \overline{P} \right\rangle$ instead of $F$. We notice that these new thresholds can be different from the ones used in Sec.~\ref{sec:SetDetection}, therefore implying that the best performing methods of Sec.~\ref{sec:SetDetection} will not necessarily yield the highest increments $\Delta$. In particular, clustering and community detection algorithms giving better community detection \textit{and} more suitable false positives will achieve the highest increments. Thus, due to the definition of $\mathscr{\hat{W}}$ [Eq.~(\ref{eqn:refineW})], the role of false positives becomes important. Furthermore, due to the use of different evaluation metrics, small changes in the optimal parameters might be necessary.% However, in general, and in spite of these facts, a substantial difference in the optimal parameters was not observed.

To illustrate the above reasoning regarding false positives consider the following example. Suppose the first items of the ranked answer to a given query $\mathring{s}_i$ are $A^0_i=\{s_j,\mathring{s}_k,s_l,s_m,\ldots\}$, where $\mathring{s}$ indicates effective (real) membership to the same cover song group. Now suppose that clustering algorithm CA1 selects songs $\mathring{s}_i$, $s_j$, and $\mathring{s}_k$ as belonging to the same cluster. In addition, suppose that clustering algorithm CA2 selects $\mathring{s}_i$, $\mathring{s}_k$, $s_l$, and $s_m$. Both clustering algorithms would have the same recall $\bar{R}$ but CA1 will have a higher precision $\bar{P}$, and therefore a higher accuracy value $F$ [Eqs.~(\ref{eqn:F}-\ref{eqn:Ri})]. Then, by Eq.~(\ref{eqn:refineW}), the refined answer for CA1 becomes $A^1_i=\{s_j,\mathring{s}_k,s_l,s_m,\ldots\}$, the same as $A^0_i$. On the other hand, the refined answer for CA2 becomes $A^2_i=\{\mathring{s}_k,s_l,s_m,s_j,\ldots\}$. This implies that, when evaluating the relative accuracy increment $\Delta$ [Eqs.~(\ref{eqn:AP}-\ref{eqn:relMAPincr})], CA2 will take a higher MAP value $\left\langle \overline{P}(\mathscr{\hat{W}}) \right\rangle$ than CA1, since $\mathring{s}_k$ is ranked before $s_j$ in $A^2_i$. Therefore, with regard to relative increments $\Delta$, and contrastingly to accuracy $F$, CA1 will not improve the result, while CA2 will.

We report the out-of-sample accuracy increments $\Delta$ for setups 2.1 to 3 in Table \ref{table:ResultsMAP}. Overall, these are between 3\% and 5\% for UPGMA, WPGMA, MO, and all PMs, with some of them reaching 6\%. We see that, in general, methods based on complex networks perform better, specially MO and PM1. We also see that the inclusion of ``noise songs'' ($N_{\mathrm{N}}=400$, setups 2.3 and 2.4) affects the performance of nearly all algorithms (with the exception of poorly performing ones).

\begin{table}[tb]
 \centering
 \begin{tabular}{lccccc}
  \hline
  Algorithm & \multicolumn{5}{c}{Setup} \\
  \cline{2-6}
            & 2.1 & 2.2 & 2.3 & 2.4 & 3\\
  \hline
  KM & 2.26 & 2.40 & 2.06 & 2.29 & \textit{n.c.} \\
  SL & 2.26 & 2.40 & 1.16 & 2.29 & 2.05 \\
  CL & 1.93 & 1.19 & 1.43 & 1.10 & 1.28 \\
  UPGMA & 5.87 & 5.22 & 3.96 & 3.49 & 4.37 \\
  WPGMA & 4.91 & 3.58 & 3.83 & 2.67 & 3.60 \\
  MO & 6.84 & 5.37 & 5.14 & 2.94 & 5.54 \\
  PM1 & 6.15 & 5.70 & 4.95 & 3.28 & 5.49 \\
  PM2 & 5.98 & 4.85 & \textit{n.c.} & \textit{n.c.} & \textit{n.c.} \\
  PM3 & 6.05 & 5.10 & 3.81 & 2.97 & 4.73 \\
  \hline
 \end{tabular}
 \caption{Relative MAP increase $\Delta$ for the considered setups (see Table~\ref{table:Setups} for the details on the different setups). Due to algorithms' complexity, some results were not computed (denoted as \textit{n.c.}).}
 \label{table:ResultsMAP}
\end{table}

A further out-of-sample test was done within the MIREX audio cover song identification contest. The MIR evaluation exchange (MIREX) is an international community-based framework for the formal evaluation of MIR systems and algorithms \citep{Downie08AST}. Among other tasks, MIREX allows for an objective assessment of the accuracy of different cover song identification algorithms. For that purpose, participants can submit their algorithms as binary executables (i.e.~as a black box, without disclosing any details), and the MIREX organizers determine and publish the algorithms' accuracies and runtimes. The underlying music collections are never published or disclosed to the participants, either before or after the contest. Therefore, participants cannot tune their algorithms to the music collections used in the evaluation process. In the editions of 2008 and 2009 we submitted the same two versions of our system and obtained the two highest accuracies achieved to date\footnote{The results for 2008 and 2009 are available from \url{http://music-ir.org/mirex/2008} and \url{http://music-ir.org/mirex/2009}, respectively. We did not participate in the 2010 edition because the MIREX evaluation dataset was kept the same and we did not have any new algorithm to submit.} \citep{Serra09MIREX}. The first version of the system (submitted to both editions) corresponded to the $Q_{\max}$ measure alone, while the second version (also submitted to both editions) comprised $Q_{\max}$ plus PM1\footnote{We just submitted PM1 because it was the only algorithm we had available at that time.} and the dissimilarity update of Eq.~(\ref{eqn:refineW}). The MAP $\left\langle \overline{P} \right\rangle$ achieved with the former was 0.66 while with the latter was 0.75. This corresponds to a relative increment $\Delta = 13.64$, which is substantially higher than the ones achieved here with our data, most probably because the setup for the MIREX task is $N_{\mathrm{C}}=30$, $C = 11$, and $N_{\mathrm{N}}=0$. Such setup might capitalize the effects that community detection can have in improving the accuracy. In particular, the techniques presented here have greater potential of increasing the final accuracies when high cardinalities are considered.

\section{The role of the original song within a cover song community \label{sec:Originals}}

%In a data clustering context, many applications exploit compact cluster descriptions \citep{Xu09BOOK, Jain99ACM}. These compact descriptions are usually given in terms of representative patterns such as the centroid or, if we want to restrict ourselves to existing elements within the cluster, the medoid. In the context of cover song networks, one might also be interested in finding a compact representative description of a cover song community. Indeed, analogously to the clustering context, the centroids and medoids of cover song communities can be determined. This way, the centroid and the medoid of a cover song community would correspond to the ``average'' realization and the ``best example'' of the underlying musical piece, respectively. In the context of cover song communities, one could consider these prototypes to be the most referential, influential, or inspirational song (e.g.~the musical piece covered by the majority of the other pieces).

From a music perception and cognition point of view, a musical work or song can be considered as a category \citep{Zbikowski02BOOK}. Categories are one of the basic devices to represent knowledge, either by humans or by machines \citep{Rogers04BOOK}. According to existing empirical evidence, some authors postulate that our brain builds categories around prototypes, which encapsulate the statistically most-prevalent category features, and against which potential category members are compared \citep{Rosch75CP}. Under this view, after the listening of several cover songs, a prototype for the underlying musical piece would be abstracted by listeners. This prototype might encapsulate features like the presence of certain motives, chord progressions, or contrasts among different musical elements. In this scenario, new items will be then judged in relation to the prototype, forming gradients of category membership \citep{Rosch75CP}.

In the context of cover song communities, we hypothesize that these gradients of category membership, in a majority of cases, might point to the original song, i.e.~the one which was firstly released. In particular we conjecture that, in one way or another, all cover songs inherit some characteristics from this ``original prototype''. This feature, combined with the fact that new versions might as well be inspired by other covers, leads us to infer that the original song occupies a \textit{central} position within a cover song community, being a referential or ``best example'' of it \citep{Serra09ISMIR}.

To evaluate this hypothesis we manually check for original versions in setup 3 and discard the sets that do not have an original, i.e.~the ones where the oldest song was not performed by the original artist.Here we make an oversimplification and assume that the most well-known (or popular) version of a song is the original one. This allows us to objectively ``mark'' our cover songs with a label stating if they are actually the original version, thus avoiding to make subjective judgments about a song's popularity with regard to its covers. Following this criteria, we find 426 originals out of 523 cover sets. Through this section, we employ the directed weighted graph defined by the asymmetric matrix $\mathscr{W}$ (Secs.~\ref{sec:NetworkBuilding} and \ref{sec:NetworkAnalysis}).

\begin{figure*}[tb]
 \centerline{\includegraphics[width=0.95\linewidth]{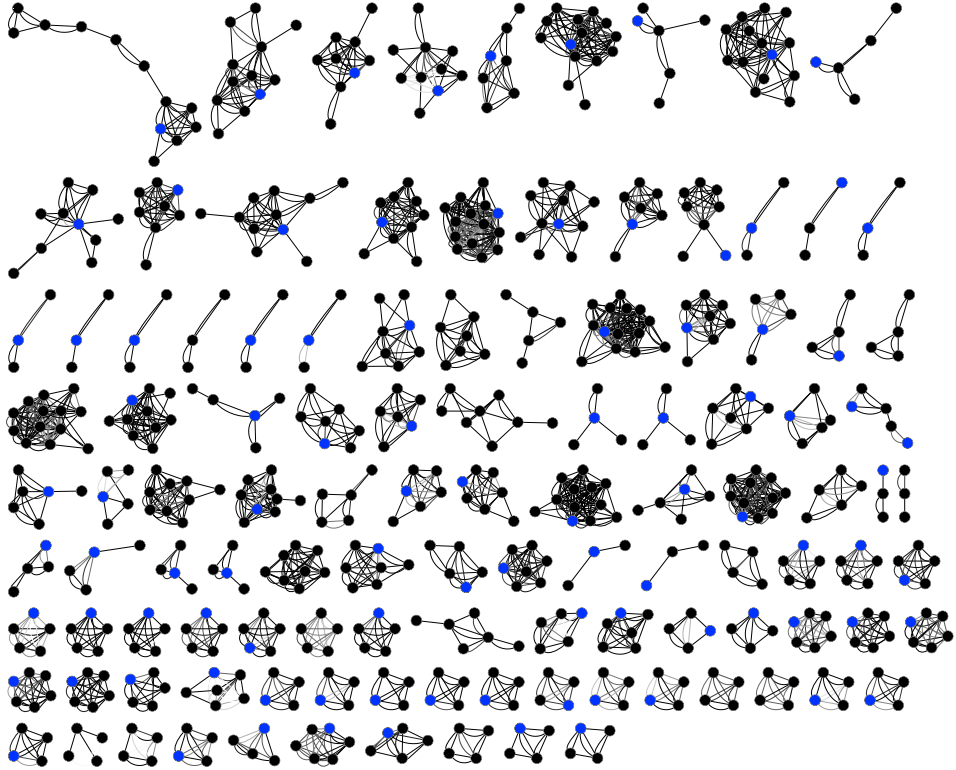}}
 \caption[sdh]{Graphical representation of the cover song network with a threshold of $0.1$. Original songs are drawn in blue, while covers are in black.}
 \label{fig:figNet2}
\end{figure*}

\begin{figure}[tb]
\centerline{\includegraphics[width=0.95\linewidth]{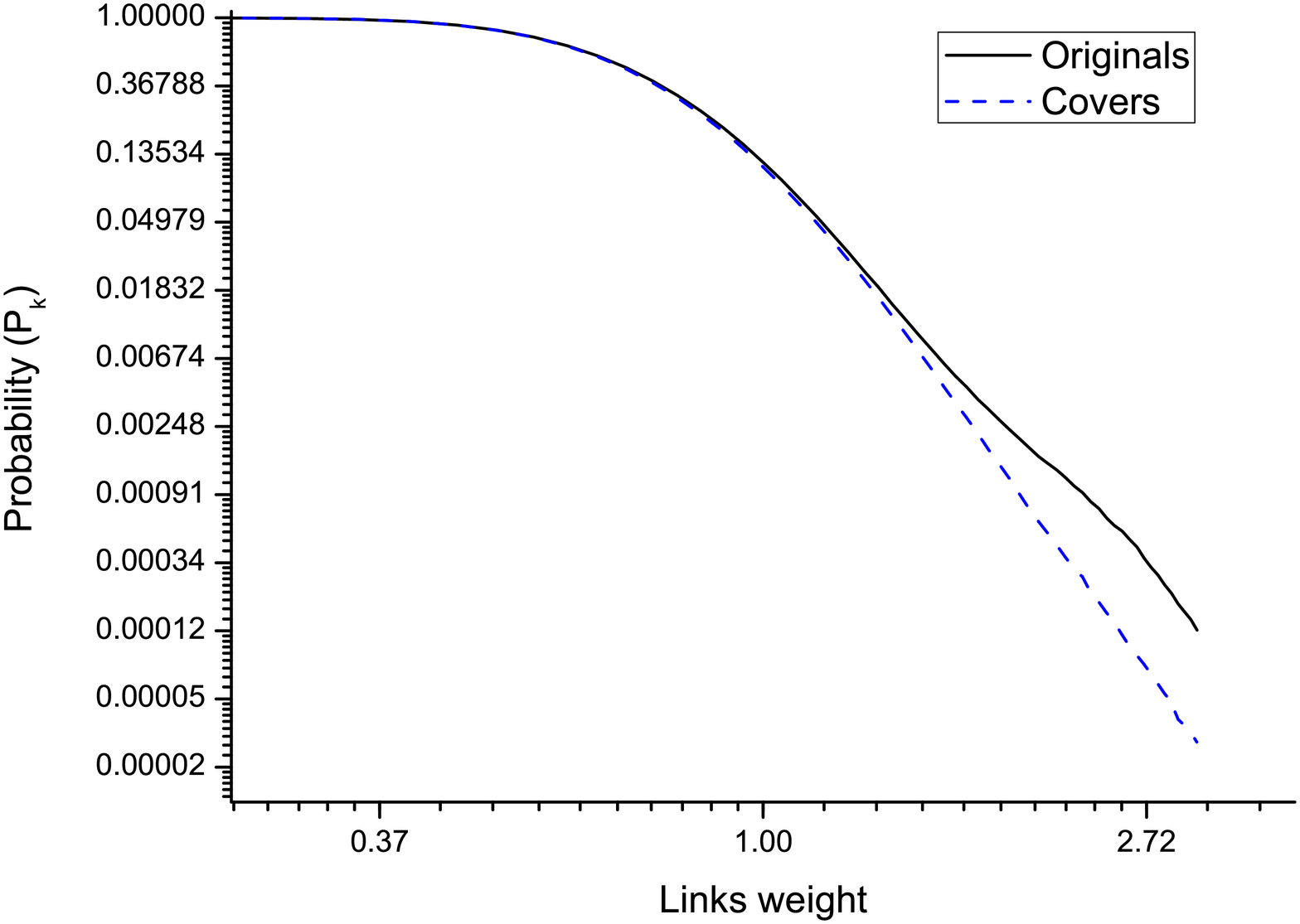}}
 \caption[sdh]{Cumulative weights distributions for links in the network, divided between links outgoing from an original song (black solid line) and from a cover song (blue dashed line) songs.}
 \label{fig:figWeights}
\end{figure}

Initial supporting evidence that the original song is central within its community is given by Figs.~\ref{fig:figNet2} and \ref{fig:figWeights}. In Fig.~\ref{fig:figNet2}, we depict the resulting network after the application of a strong threshold (only using $w_{i,j}\leq 0.1$). We see that communities are well defined and also that many of the original songs are usually ``the center'' of their communities. In Fig.~\ref{fig:figWeights}, two cumulative distributions have been calculated: one for the weights of links exiting an original song (performed by the original artist, black solid line), and one for links exiting covers (performed by the original artist or another one after the original recording was made, blue dashed line). The plot of these cumulative distributions indicates that original songs tend to be connected to other nodes through links with smaller weights, that is, lower dissimilarities.

To evaluate the aforementioned hypothesis in a more formal way, we propose a study of the ability to automatically detect the original version within a community of covers. To this extent, we consider an ``ideal'' community detection algorithm (i.e.~an algorithm detecting cover song communities with no false positives and no false negatives) and propose two different methods. These methods are based on the structure of weights of the obtained sub-network after the ideal community detection algorithm has been applied.

\begin{description}

\item[Closeness centrality] This algorithm estimates the centrality \citep{Boccaletti06PR, Barrat04PNAS} of a node by calculating the mean path length between that node, and any other node in the sub-network. Note that the sub-network is fully connected, as no threshold has been applied in this phase. Therefore, the shortest path is usually the direct one. Mathematically, let $\mathscr{W}^{(k)}$ be the sub-network containing the $k$-th cover song community. Then the index $l$ of the original (or prototype) song $s^{(k)}_l$ of the $k$-th community corresponds to
\begin{equation}
\displaystyle
l = \underset{1\leq i \leq C^{(k)}}{\operatorname{arg\,min}} \left\lbrace \sum_{\substack{j=1 \\ j\neq i}}^{C^{(k)}} w^{(k)}_{i,j} \right\rbrace ,
\label{eqn:Originals}
\end{equation}
where $C^{(k)}$ is the cardinality of the $k$-th cover song community. Notice that a similar methodology is employed in the clustering context to infer the medoid of a cluster \citep{Xu09BOOK, Jain99ACM}.

\item[MST centrality] In this second algorithm we reinforce the role of central nodes. First, we calculate the minimum spanning tree (MST) for the sub-network under analysis. After that, we apply the previously described closeness centrality [Eq.~(\ref{eqn:Originals})] to the resulting graph.

\end{description}

The results in Table~\ref{table:Originals} show the percentage of hits and misses for the detection of original songs in dependence of the cardinality of the considered cover song community. We report results for $C$ between 2 and 7 (the cardinalities for which our music collection has a representative number of communities $N_{\mathrm{C}}$). The percentage of hits and misses can be compared to the null hypothesis of randomly selecting one song in the community.

\begin{table*}[tb]
 \centering
 \begin{tabular}{lcccccc}
  \hline
  Algorithm & \multicolumn{6}{c}{$C$} \\
  \cline{2-7}
            & 2 & 3 & 4 & 5 & 6 & 7 \\
  \hline
  Closeness centrality & 59.4$^{\ast\ast}$ & 53.6$^{\ast\ast}$ & 43.1$^\ast$ & 60.5$^{\ast\ast}$ & 48.0$^{\ast\ast}$ & 27.2 \\
  MST centrality & 50.0 & 52.4$^{\ast\ast}$ & 60.7$^{\ast\ast}$ & 52.6$^{\ast\ast}$ & 48.0$^{\ast\ast}$ & 63.6$^{\ast\ast}$ \\
  \hline
  Null hypothesis & 50.0 & 33.3 & 25.0 & 20.0 & 16.7 & 14.3 \\
  \hline
  $N_{\mathrm{C}}$ & 190 & 82 & 51 & 38 & 25 & 11 \\
  \hline
 \end{tabular}
 \caption{Percentage of hits and misses for the original song detection task depending on the cardinality $C$ of the cover song communities. The $^\ast$ and $^{\ast\ast}$ symbols denote statistical significance at $p<0.05$ and $p<0.01$, respectively. The last line shows $N_{\mathrm{C}}$, i.e.~the number of communities for each cardinality.}
 \label{table:Originals}
\end{table*}

We observe that, in general, accuracies are around 50\% and, in some cases, they reach values of 60\%. An accuracy of exactly 50\% is obtained with $C=2$ by both the null hypothesis and the MST centrality algorithm. This is because the MST is defined undirected, and there is no way to discriminate the original song in a sub-network of two nodes. As soon as $C>2$, accuracies become greater than the null hypothesis and statistical significance arises. Statistical significance is assessed with the binomial test \citep{Kvam07BOOK}.

With this experiment we show that the original song tends to occupy a central position within its group and, therefore, that a measure of centrality can be used to discriminate it from a group of covers. The same concepts of centrality may be valid for alternative dissimilarity measures representing musical aspects such as timbre, rhythm, or structure \citep[c.f.][]{Downie08AST, Casey08IEEE}. Thus, one could think of incorporating information from these other aspects of the audio content in order to improve the accuracy of the task. A more complicated, if not impossible, task would be to detect the original song in a pairwise basis. To this extent, works on modeling court decisions like the ones from \citet{Mullensiefen09MS} come closer. In general, for detecting original songs, information coming from the audio content alone may be insufficient. Essential temporal aspects (in a historical sense) are absent in such information and, for incorporating them, we should gather data from cultural and editorial sources. This goes without saying that, probably, high accuracies are unreachable and, more importantly, that the concept of originality is a very particular one, placed in a specific cultural context and epoch. Indeed, the digital revolution of the last years is beginning to question such a concept \citep{Fitzpatrick09CM}.

\section{Conclusions \label{sec:Conclusions}}

In this article we built and analyzed a musical network reflecting cover song communities, where nodes corresponded to different audio recordings and links between them represented a measure of resemblance between their musical content. In addition, we analyzed the possibility of using such a network to apply different community detection algorithms to detect coherent groups of cover songs. Three versions of such algorithms were proposed. These algorithms achieved comparable accuracies when compared to existing state-of-the-art methods, with similar or even faster computation times. Furthermore, we provide evidence that the knowledge acquired through community detection is valuable in improving the raw results of a query-based cover song identification system. Finally, we discussed a particular outcome from considering cover song communities, namely the analysis of the role of the original song within its covers. We showed that the original song tends to occupy a central position within its group and, therefore, that a measure of centrality can be used to discriminate original from cover songs when the sub-network of these communities is considered. To the best of authors' knowledge, the present work is the first attempt done in this direction.

In the light of these results, complex networks stand as a promising research line within the specific task of cover detection; but, at the same time, the proposed approach can be applied to any query-by-example IR system \citep{BaezaYates99BOOK, Manning08BOOK}, and especially to other query-by-example MIR systems \citep{Downie08AST, Casey08IEEE}.

\section*{Acknowledgments}

The authors thank Justin Salamon for his review and proofreading. This work has been supported by the following projects: Classical
Planet (TSI-070100- 2009-407; MITYC) and DRIMS (TIN2009-14247-C02-01; MICINN).

\section*{References}

\bibliographystyle{elsarticle-harv}
%\bibliography{jserra}

\begin{thebibliography}{30}
\expandafter\ifx\csname natexlab\endcsname\relax\def\natexlab#1{#1}\fi
\expandafter\ifx\csname url\endcsname\relax
  \def\url#1{\texttt{#1}}\fi
\expandafter\ifx\csname urlprefix\endcsname\relax\def\urlprefix{URL }\fi

\bibitem[{Baeza-Yates and Ribeiro-Neto(1999)}]{BaezaYates99BOOK}
Baeza-Yates, R., Ribeiro-Neto, B., 1999. Modern information retrieval. ACM
  Press, New York, USA.

\bibitem[{Barrat et~al.(2004)Barrat, Barth{\'e}lemy, Pastor-Satorras, and
  Vespignani}]{Barrat04PNAS}
Barrat, A., Barth{\'e}lemy, M., Pastor-Satorras, R., Vespignani, A., 2004. The
  architecture of complex weighted networks. Proc. of the National Academy of
  Sciences 101, 3747.

\bibitem[{Blondel et~al.(2008)Blondel, Guillaume, Lambiotte, and
  Lefebvre}]{Blondel08JSM}
Blondel, V.~D., Guillaume, J.~L., Lambiotte, R., Lefebvre, E., 2008. Fast
  unfolding of communities in large networks. Journal of Statistical Mechanics
  10, 10008.

\bibitem[{Boccaletti et~al.(2006)Boccaletti, Latora, Moreno, Chavez, and
  Hwang}]{Boccaletti06PR}
Boccaletti, S., Latora, V., Moreno, Y., Chavez, M., Hwang, D.-U., 2006. Complex
  networks: structure and dynamics. Physics Reports 424~(4), 175--308.

\bibitem[{Buld{\'u} et~al.(2007)Buld{\'u}, Cano, Koppenberger, Almendral, and
  Boccaletti}]{Buldu07NJP}
Buld{\'u}, J.~M., Cano, P., Koppenberger, M., Almendral, J., Boccaletti, S.,
  2007. The complex network of musical tastes. New Journal of Physics 9, 172.

\bibitem[{Cano et~al.(2006)Cano, Celma, Koppenberger, and Buld{\'u}}]{Cano06C}
Cano, P., Celma, O., Koppenberger, M., Buld{\'u}, J.~M., 2006. Topology of
  music recommendation networks. Chaos: an Interdisciplinary Journal of
  Nonlinear Science 16~(1), 013107.

\bibitem[{Casey et~al.(2008)Casey, Veltkamp, Goto, Leman, Rhodes, and
  Slaney}]{Casey08IEEE}
Casey, M., Veltkamp, R.~C., Goto, M., Leman, M., Rhodes, C., Slaney, M., 2008.
  Content-based music information retrieval: current directions and future
  challenges. Proceedings of the IEEE 96~(4), 668--696.

\bibitem[{Costa et~al.(2008)Costa, Oliveira, Travieso, Rodrigues, Villas~Boas,
  Antiqueira, Viana, and Correa~da Rocha}]{Costa08preprint}
Costa, L. d.~F., Oliveira, O.~N., Travieso, G., Rodrigues, F.~A., Villas~Boas,
  P.~R., Antiqueira, L., Viana, M.~P., Correa~da Rocha, L.~E., 2008. Analyzing
  and modeling real-world phenomena with complex networks: a survey of
  applicationsWorking manuscript, ar{X}iv:0711.3199v2. Available online:
  \url{http://arxiv.org/abs/0711.3199}.

\bibitem[{Danon et~al.(2005)Danon, D{\'{\i}}az-Aguilera, Duch, and
  Arenas}]{Danon05JSM}
Danon, L., D{\'{\i}}az-Aguilera, A., Duch, J., Arenas, A., 2005. Comparing
  community structure identification. Journal of Statistical Mechanics 9,
  09008.

\bibitem[{Downie(2008)}]{Downie08AST}
Downie, J.~S., 2008. The music information retrieval evaluation exchange
  (2005--2007): a window into music information retrieval research. Acoustical
  Science and Technology 29~(4), 247--255.

\bibitem[{Eckmann et~al.(1987)Eckmann, Kamphorst, and Ruelle}]{Eckmann87EPL}
Eckmann, J.~P., Kamphorst, S.~O., Ruelle, D., 1987. Recurrence plots of
  dynamical systems. Europhysics Letters 5, 973--977.

\bibitem[{Fitzpatrick(2009)}]{Fitzpatrick09CM}
Fitzpatrick, K., 2009. The digital future of authorship: rethinking originality.
 Culture Machine, vol. 12, 6.

\bibitem[{Fortunato and Castellano(2009)}]{Fortunato09BOOKCHAP}
Fortunato, S., Castellano, C., 2009. Community structure in graphs. In: Meyers,
  R.~A. (Ed.), Encyclopedia of complexity and system science. Springer, Berlin,
  Germany, pp. 1141--1163.

\bibitem[{G{\'o}mez(2006)}]{Gomez06THESIS}
G{\'o}mez, E., 2006. Tonal description of music audio signals. Ph.D. thesis,
  Universitat Pompeu Fabra, Barcelona, Spain, {A}vailable online:
  \url{http://mtg.upf.edu/node/472}.

\bibitem[{Jain et~al.(1999)Jain, Murty, and Flynn}]{Jain99ACM}
Jain, A.~K., Murty, M.~N., Flynn, P.~J., 1999. Data clustering: a review. ACM
  Computing Surveys 31~(3), 264--323.

\bibitem[{Kantz and Schreiber(2004)}]{Kantz04BOOK}
Kantz, H., Schreiber, T., 2004. Nonlinear time series analysis, 2nd Edition.
  Cambridge University Press, Cambridge, UK.

\bibitem[{Kvam and Vidakovic(2007)}]{Kvam07BOOK}
Kvam, P.~H., Vidakovic, B., 2007. Nonparametric statistics with applications to
  science and engineering. John Wiley and Sons, Hoboken, USA.

\bibitem[{Lagrange and Serr{\`a}(2010)}]{Lagrange10ISMIR}
Lagrange, M., Serr{\`a}, J., 2010. Unsupervised accuracy improvement for cover
 song detection using spectral connectivity network. Proc. of the Int. Soc. for
 Music Information Retrieval, pp. 595--600.

\bibitem[{Latora and Marchiori(2001)}]{Latora01PRL}
Latora, V., Marchiori, M., 2001. Efficient behavior of small-world networks.
  Physical Review Letters 87, 198701.

\bibitem[{Manning et~al.(2008)Manning, Prabhakar, and Schutze}]{Manning08BOOK}
Manning, C.~D., Prabhakar, R., Schutze, H., 2008. An introduction to
  information retrieval. Cambridge University Press, Cambridge, UK.

\bibitem[{Marwan et~al.(2007)Marwan, Romano, Thiel, and Kurths}]{Marwan07PR}
Marwan, N., Romano, M.~C., Thiel, M., Kurths, J., 2007. Recurrence plots for
  the analysis of complex systems. Physics Reports 438~(5), 237--329.

\bibitem[{M{\"u}llensiefen and Pendzich(2009)}]{Mullensiefen09MS}
M{\"u}llensiefen, D., Pendzich, M., 2009. Court decisions on music plagiarism and
 the predictive value of similarity algorithms. Musicae Scientiae, Discussion
 Forum 4B, 207--238.

\bibitem[{Resnick and Varian(1997)}]{Resnick97CACM}
Resnick, P., Varian, H.~L., 1997. Recommender systems. Communications of the
  ACM 40~(3), 56--58.

\bibitem[{Rogers and McClelland(2004)}]{Rogers04BOOK}
Rogers, T.~T., McClelland, J.~L., 2004. Semantic cognition: a parallel
  distributed processing approach. MIT Press, Cambridge, USA.

\bibitem[{Rosch and Mervis(1975)}]{Rosch75CP}
Rosch, E., Mervis, C., 1975. Family resemblances: studies in the internal
  structure of categories. Cognitive Psychology 7, 573--605.

\bibitem[{Sahoo et~al.(2006)Sahoo, Callan, Krishnan, Duncan, and
  Padman}]{Sahoo06IKM}
Sahoo, N., Callan, J., Krishnan, R., Duncan, G., Padman, R., 2006. Incremental
  hierarchical clustering of text documents. In: Proc. of the ACM Int. Conf. on
  Information and Knowledge Management. pp. 357--366.

\bibitem[{Serr{\`a} et~al.(2010)Serr{\`a}, G{\'o}mez, and
  Herrera}]{Serra10BOOKCHAP}
Serr{\`a}, J., G{\'o}mez, E., Herrera, P., 2010. Audio cover song
  identification and similarity: background, approaches, evaluation, and
  beyond. In: Ras, Z.~W., Wieczorkowska, A.~A. (Eds.), Advances in Music
  Information Retrieval. Vol.~16 of Studies in Computational Intelligence.
  Springer, Berlin, Germany, Ch.~14, pp. 307--332.

\bibitem[{Serr{\`a} et~al.(2009{\natexlab{a}})Serr{\`a}, Serra, and
  Andrzejak}]{Serra09NJP}
Serr{\`a}, J., Serra, X., Andrzejak, R.~G., 2009{\natexlab{a}}. Cross
  recurrence quantification for cover song identification. New Journal of
  Physics 11, 093017.

\bibitem[{Serr{\`a} et~al.(2009{\natexlab{c}})Serr{\`a}, Zanin, and
  Andrzejak}]{Serra09MIREX}
Serr{\`a}, J., Zanin, M., Andrzejak, R.~G., 2009{\natexlab{c}}. Cover song
  retrieval by cross recurrence quantification and unsupervised set detection.
  Music Information Retrieval Evaluation eXchange (MIREX) extended abstract.

\bibitem[{Serr{\`a} et~al.(2009{\natexlab{b}})Serr{\`a}, Zanin, Laurier and
  Sordo}]{Serra09ISMIR}
Serr{\`a}, J., Zanin, M., Laurier, C., Sordo, M., 2009{\natexlab{b}}. Unsupervised
 detection of cover song sets: accuracy improvement and original identification.
 Proc. of the Int. Soc. for Music Information Retrieval, pp. 225--230.

\bibitem[{Teitelbaum et~al.(2008)Teitelbaum, Balenzuela, Cano, and
  Buld{\'u}}]{Teitelbaum08C}
Teitelbaum, T., Balenzuela, P., Cano, P., Buld{\'u}, J.~M., 2008. Community
  structures and role detection in music networks. Chaos: an Interdisciplinary
  Journal of Nonlinear Science 18~(4), 043105.

\bibitem[{Xu and Wunsch~II(2009)}]{Xu09BOOK}
Xu, R., Wunsch~II, D.~C., 2009. Clustering. IEEE Press, Piscataway, USA.

\bibitem[{Zbikowski(2002)}]{Zbikowski02BOOK}
Zbikowski, L.~M., 2002. Conceptualizing music: cognitive structure, theory, and
  analysis. Oxford University Press, Oxford, UK.

\bibitem[{Zhao and Karypis(2002)}]{Zhao02KDD}
Zhao, Y., Karypis, G., 2002. Evaluation of hierarchical clustering algorithms
  for document datasets. In: Proc. of the Conf. on Knowledge Discovery in Data
  (KDD). pp. 515--524.

\end{thebibliography}

\end{document}